\documentstyle[pre,aps,eqsecnum, array, epsfig,multicol ]{revtex}

\begin{document}
\draft
\title{Numerical studies of domains and bubbles of Langmuir monolayers}

\author{Kok-Kiong Loh and Joseph Rudnick}
\address{Department of Physics, UCLA \\
Los Angeles, California 90095-1547}
\date{\today}
\maketitle

\begin{abstract}
A numerical algorithm based on the finite element methods has been
developed to accurately determine the shape of the boundary of a
domain containing ``boojum'' textures.  Within the context of the
simple model we adopt, the effects of both bulk elastic anisotropy and
line-tension anisotropy on the domain boundary can be examined.  It is
found that line-tension anisotropy must be present in order to account
for domains with protruding features.  Both elastic anisotropy and
anisotropic line-tension can result in domains with indentations.  The
numerical algorithm has been extended to investigate the problem of a
bubble in extended region ordered phase.
\end{abstract}

\begin{multicols}{2}

\section{Introduction}

A Langmuir monolayer is a single molecular layer of insoluble
surfactant molecules spread on the air/water interface.  The
surfactants are typically amphiphilic molecules with a hydrophilic
headgroup and a hydrophobic tail.  Each of the individual molecules
has internal degrees of freedom, namely the tilt and the tilt azimuth.
Such a system exhibits complex phase structure\cite{phases}.  The
``tilted'' phases have uniform tilt and possess mesoscopic ordering in
the tilt azimuth.  The structure of the tilt azimuth is typically
observed as a variation in the light intensity under a polarized light
microscope.  The tilt azimuth organization is referred to as the
texture.  Various classes of the texture have been observed, such as
the stripes in the bulk\cite{stripe}, the star
configuration\cite{star,FiscBru} and ``boojums''\cite{RivMeu} in the
domains of the tilted phase, when it coexists with an isotropic phase.
The term Boojum refers to a class of textures that has a tilt azimuth
distribution which resembles the structure of the orbital angular
momentum in a superfluid $^3$He droplet\cite{Mermin}.  The ``boojum''
texture, in which the tilt azimuth is distributed continuously without
singularity, will be the subject of this report.  Domains observed to
contain a boojum texture are not circular in
shape\cite{RivMeu,FangTeer}.  In addition, micron-size bubbles, which
are regions of isotropic phase surrounded by a ``tilted'' phase, have
been found to have non-circular shapes \cite{FangTeer}.  The local
tilt azimuth in the ``tilted'' phase around the bubble exhibits a
non-trivial structure, which has been termed an ``inverse boojum''.
The relationship between experimentally observed textures and the
underlying structure of the ordered phase has attracted attention in
the literature recently.  In particular, the ``boojum'' texture was
first discussed by Mermin in the context of orbital angular momental
distribution in superfluid $^3$He droplet \cite{Mermin}.  Similar
texture of has been found and discussed in the liquid crystal
films\cite{LanSeth,Kraus}.  An extensive discussion on the various
classes of textures in the Langmuir monolayers can be found in
Ref.~\cite{FiscBru}.

The problem of the equilibrium shape of, and the texture contained in
domains in a Langmuir monolayer has been investigated by Rudnick and
Bruinsma \cite{RudBru} who varied both the texture and the boundary
analytically in a perturbative manner.  It was discovered that a
non-circular boundary represents the equilibrium shape of a domain
only when there is bulk or line-tension
anisotropy\cite{RivMeu,RudBru}.  The equilibrium domain boundary was
derived as a function of line-tension anisotropy.  Galatola and
Fournier \cite{GalaFour} obtained numerically, in a fixed background
texture, the equilibrium boundary when both elastic and line-tension
anisotropies are present.  Rivi\`{e}re and Meunier \cite{RivMeu} have
attempted to explain the observed non-circular domains in terms of
elastic anisotropy.  Evidence of bubbles with a non-circular boundary
and an ``inverse boojum'' has been reported, and a qualitative
theoretical discussion of the equilibrium shape and texture associated
with the bubbles can be found in Ref.~\cite{FangTeer}. In the spirit of
Ref.~\cite{RudBru}, the authors have analyzed in Ref.~\cite{pertb} the
equilibrium texture and boundary shape combinations perturbatively to
the first order in both the bulk elastic and line-tension
anisotropies.  The approach describes the infinitesimal response of
the texture and the boundary to anisotropic parameters.  However, when
the correction is large enough to be observed, the validity of first
order perturbative calculations becomes questionable.  The extension
of the perturbative approach to include higher order corrections is
algebraically formidable.  If one is to go beyond first order effects,
the use of numerical techniques in this problem is inevitable.

The major challenges in this problem are, first, the evaluation of a
2D texture with a boundary condition on the boundary, which is,
itself, variable.  Secondly, not only must the texture be evaluated
with high accuracy, but a precise determinations of the derivatives of
the texture on the boundary is also crucial to the computation of the
boundary shape.  The authors have developed a numerical algorithm
based on the finite element method (FEM) with adaptive mesh refinement
\cite{FEM} for the evaluation of a 2D texture and its derivatives.  
The boundary corrections can then be computed using the Runge-Kutta 
methods\cite{recipes}. Implementation of the numerical method reveals 
various classes of domain shapes ranging from those with indentations 
to those with protruding features and, additionally, of cigar-shaped 
domains.  The effects of bulk elastic anisotropy have also been 
examined.  These studies lead us to the conclusion that, as least 
for those domain shapes observed to date, it is more likely that the 
line-tension anisotropy is responsible for non-circular domains.  A brief 
account of the study described above has appeared in an earlier 
publication\cite{numerics}.  The numerical results also confirm that the
qualitative conclusion to be drawn from the perturbative treatment are
preserved up to large anisotropic parameter.

In this report, we describe in detail the implementation of the
numerical methods that lead us to the results reported in
Ref~\cite{numerics}.  The extension of the algorithm to allow for
computation in the case of bubble has also been examined.  It is
verified numerically that bubbles acquire a non-trivial boundary shape
when only the first term in the Fourier expansion of the line tension
is present.  This result contrasts with what is known to be true in
the case of domain, which remains circular in the presence of this
low-order line-tension anisotropy\cite{RudBru}.  With the use of our
numerical algorithm, we are able to examine the effects of the bulk
elastic anisotropy on the shape of the bubble and on the texture that
surrounds it.  We find that bulk elastic anisotropy significantly
affects the texture in the condensed phase around the bubble while
leaving the boundary nearly unmodified.

The organization of this paper is as follows.
Section~\ref{sec:numerics} contains the details of the computational
scheme for the evaluation of the equilibrium textural and boundary
configuration for domains.  The discussion covers the derivation of
the simplest variational formulation of the finite element method in
our specific application, the Runge-Kutta method and the combined
algorithm.  In Sec.~\ref{sec:domains}, results for the domain are
examined.  Section~\ref{sec:bubbles} describes the extension of the
numerical algorithm to the problem of bubbles.  An examination of the
results of the perturbative treatment follows.  New results on the
effect of the bulk elastic anisotropy on the textures around the
bubbles are discussed.  Finally, Sec.~\ref{sec:conclusions} contains
concluding remarks and discusses possible future extensions of the
numerical methods discussed in this report.

\section{The Numerical Algorithm}
\label{sec:numerics}
The model that we adopt for the Langmuir monolayer is a simple elastic
model of an ordered media associated with $XY$-like order
parameter---a 2 dimensional unit vector $\hat{c}(x, y)$, which can be
parameterized as $\hat{x}\cos\Theta(x, y)+\hat{y}\sin\Theta(x, y)$
~\cite{pertb}.  The quantities $\hat{x}$ and $\hat{y}$ are unit
vectors in a Cartesian coordinate system, and $\Theta(x, y)$ is the
angle between $\hat{c}$ and the x-axis.  The energy of the system
contains contributions from the boundary, $\Gamma$, in addition to the
bulk, $\Omega$.  The most general form of the elastic
energy\cite{FiscBru,LanSeth} for such a system with in-plane
reflection symmetry (an achiral system) can be written as
\begin{eqnarray}
H[\Theta]=\int_\Omega {\cal H}_b dA+\oint_\Gamma\sigma(\vartheta-\Theta)ds,
\label{sysenr}
\end{eqnarray}
where
\begin{eqnarray}
{\cal H}_b&=&\frac{K_s}{2}|\nabla\cdot\hat{c}|^2+\frac{K_b}{2}|\nabla\times
\hat{c}|^2,\\
\sigma(\phi)&=&\sigma_0+\sum_{n=1}a_n\cos n\phi,\label{linetension}
\end{eqnarray}
$K_s$, $K_b$ are respectively the splay and bend elastic moduli, and
$\vartheta$ is the angle between the outward normal $\hat{n}$ of
$\Gamma$ and the x-axis.  The setup of the computations is shown in
Fig.~\ref{setup}.  In terms of the average Frank modulus $\kappa$ and
the coefficient of elastic anisotropy, $b$ where $2\kappa\equiv
K_b+K_s$ and $2\kappa b \equiv K_s-K_b$, the extrema of the elastic
\vadjust{\vskip 0.1in
\narrowtext
\begin{figure}
\centerline{\epsfig{file=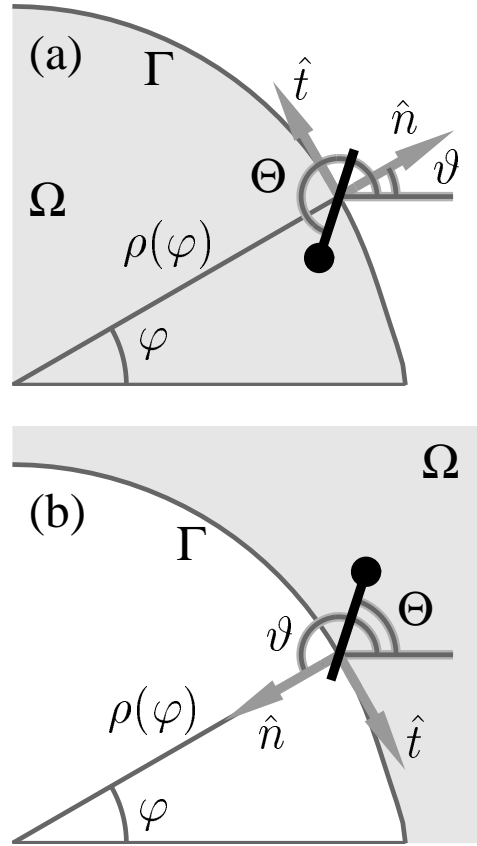, width=3in}}
\caption{The geometry of the calculations for (a) domains and (b) bubbles
in plane-polar coordinates where the boundary $\Gamma$ is parameterized by
$\rho(\varphi)$.  The gray area is the bulk designated by $\Omega$.
$\hat{n}$ and $\hat{t}$ are
the outward normal and the tangent, respectively. $\Theta$ is the angle
between the $\hat{c}$-director and the x-axis and $\vartheta$ is the angle
between the outward normal of the boundary and the x-axis.  }
\label{setup}
\end{figure}
\vskip 0.05in}
energy Eq.~(\ref{sysenr}) occurs when $\Theta(x, y)$ and the bounding
curve $\Gamma$ satisfy their respective equilibrium conditions.  The
extremum equations for $\Theta(x, y)$ are
\begin{eqnarray}
-\nabla^2\Theta + b\left[\left(\Theta_{xx}-\Theta_{yy}\right)\cos 2
\Theta +2\Theta_{xy} \sin 2 \Theta \right.\nonumber\\
\left.+ \left(-\Theta_x^2+\Theta_y^2\right)\sin
2\Theta+2\Theta_x\Theta_y \cos 2 \Theta\right]&&= 0, \label{mbulk}
\end{eqnarray}
in $\Omega$ and
\begin{eqnarray}
\kappa\Theta_n\left[1 - b \cos 2 (\vartheta-\Theta )\right]+&&\nonumber\\
\kappa b\Theta_t\sin 2(\vartheta-\Theta)-
\sigma^{\prime}(\vartheta-\Theta)&&=0, \label{mbc}
\end{eqnarray}
along $\Gamma$, where $\Theta_n=\hat{n}\cdot\nabla\Theta$,
$\Theta_t=\hat{t}\cdot\nabla\Theta$, $\hat{t}$ being the tangential
vector.  The extremum equation for the bounding curve $\Gamma$, in
terms of $\Theta_n$, $\Theta_t$ and $d\vartheta/ds$, is
\begin{eqnarray}
{\cal
H}_b-\sigma^{\prime}(\vartheta-\Theta)\Theta_n-\sigma^{\prime\prime}
(\vartheta-\Theta)\Theta_t && \nonumber \\
+\left[\sigma(\vartheta-\Theta)+\sigma^{\prime\prime}(\vartheta-\Theta)\right]
\frac{d\vartheta}{ds} + \lambda&&=0,\label{mGamma}
\end{eqnarray}
where $ds$ is the length element of $\Gamma$ traversing in the
positive direction of $\Omega$ and $\lambda$ is a Lagrange multiplier
that enforces the condition of constant enclosed area.

The equations for both $\Theta$ and $\Gamma$ are complex and highly
non-linear.  Closed form analytic solution of the extremum equations
is almost impossible.  Attempts have been made to solve the
simultaneous equation perturbatively to first order in the elastic and
line-tension anistropies~\cite{pertb}.  When the corrections to the
boundaries are large enough to be observable, it is not expected that
the results are accurate and high order corrections have to be taken
into account.  However, these attempts provide us with insight with
regard to the infinitesimal response of the boundary to the
anisotropies under investigation.  In the work to be described below,
we analyze the equations numerically in order to further explore the
implications of the simple model Eq.~(\ref{sysenr}) for a larger range
of the anisotropic parameters.  We retain coefficients up to $a_2$ in
the expansion of the line tension in our analysis, i.e.
$\sigma(\phi)= \sigma_0+a_1(\cos\phi+\gamma\cos 2\phi)$, where the
quantity $\gamma\equiv a_2/a_1$ is defined for convenience.  We remark
that the analysis will be based on the exact ``boojum'' texture with
circular domain when $\gamma=b=0$.  The boundary will be computed in
terms of the corrections to the circular boundary.  The discussions
will be restricted to those domains with boundaries $\Gamma$ for which
the distance from each points on the curves to the origin
$\mbox{e}^{k(\varphi)}$ is a single-valued function of the polar angle
$\varphi$.

The numerical algorithm consists of two parts: in the first part, one
evaluates the texture $\Theta$ using an assumed boundary $\Gamma$,
and, in the second part, one computes $\Gamma$ using a fixed $\Theta$.
Simultaneous equilibrium conditions for $\Gamma$ and $\Theta$ are
achieved when a set of predefined self-consistent criteria is met.  It
is evident from the form of Eq.~(\ref{mGamma}) that accurate
determinations of $\Theta$ and its derivatives are the key factors in
the solution of the problem.  The requirement that the assumed
$\Gamma$ in the first part of the algorithm be an arbitrary curve
rules out finite difference methods, and militates in favor of finite
element methods (FEM).  A key feature of the FEM is flexibility in the
choice the set of points at which the functional values are to be
evaluated, including those on the boundary of the region of interest.
This feature is exactly what is needed in our problem, because of the
non-trivial geometry of the boundary.  One of the simplest
constructions of the FEM in 2 dimension is described as
follows\cite{FEM}.  We first approximate $\Gamma$ by a polygonal
curve, then subdivide $\Omega$ into a set of non-overlapping
triangles.  No vertex of one triangle lies on the edge of another in
the set.  The edges of the set of triangle forms a mesh that covers
$\Omega$.  The process of creating this set of triangles is called
mesh generation.  The resulting set of triangles is referred to as the
triangulation of $\Omega$.  Functions are defined by their values on
the vertices of the triangles in the triangulation.  The value of a
function within a triangle is obtained by interpolation using the
values on the vertices.  Integration over $\Omega$ is the sum of
integrations over the triangles which can generally be trivially
evaluated.  We have now projected our problem, originally on an
infinite dimensional space onto a N dimensional space, where N is the
number of vertices in the triangulation of $\Omega$.  We may write
${\bf \Theta}\equiv(\Theta_i)$, $i=1,\cdots,{\mbox N}$ and
\begin{equation}
\Theta(x, y)=\sum_{i=1}^{N}\Theta_i \varphi_i(x, y),\label{ThetaTria}
\end{equation}
where $\varphi_i(x, y)$ is a set of basis functions of the N
dimensional space.  These $\varphi_{i}$'s should not be confused with
the polar angle which is denoted by the symbol $\varphi$ without a
subscript.  The discrete version of the elastic energy functional Eq.
(\ref{sysenr}) is a function of N variables $\Theta_i$ and it can be
rewritten in terms of $\kappa$ and $b$ as
\begin{eqnarray}
H({\bf
\Theta})&=&\frac{\kappa}{2}\int_\Omega\left\{|\nabla\Theta|^2+b\left[\left
(-\Theta_x^2+\Theta_y^2\right)\cos
2\Theta-\right.\right.\nonumber\\
&&\left.\left.2\Theta_x\Theta_y\sin 2\Theta\right]\right\}\;dA+
\oint_\Gamma\sigma(\vartheta-\Theta)\;ds,
\end{eqnarray}
where $\Theta_x=\sum \Theta_i\varphi_{ix}$,
$\Theta_y=\sum\Theta_i\varphi_{iy}$,
$\varphi_{ix}\equiv\partial\varphi_i/\partial x$ and
$\varphi_{iy}\equiv\partial\varphi_i/\partial y$.  The equilibrium
condition becomes
\begin{equation}
\frac{\partial H({\bf \Theta})}{\partial \Theta_i}=0,\;\;\; i=1,
\cdots, {\mbox N},
\label{FEM}
\end{equation}
which is a discretized version of Eqs.  (\ref{mbulk}) and (\ref{mbc}).
The set of above equations is not linear.  However, if we write them
in the form of ${\bf A(\Theta)}\cdot{\bf \Theta}={\bf b(\Theta)}$
where ${\bf A(\Theta)}$ is an ${\mbox N}\times{\mbox N}$ matrix, ${\bf
b(\Theta)}$ and ${\bf \Theta}$ are $1\times{\mbox N}$ column matrix as
shown below,
\begin{eqnarray}
A_{ij}({\bf \Theta})&=&\kappa\int_\Omega[\varphi_{ix}\varphi_{jx}
(1-b\cos 2\Theta)+\nonumber\\
&&\varphi_{iy}\varphi_{jy}(1+b\cos 2\Theta)+\nonumber\\
&&b(\varphi_{ix}\varphi_{jy}+\varphi_{iy}\varphi_{jx})\sin 2\Theta]dA
\label{MatrixA}\\
b_i(\bf \Theta)&=&\kappa b\int_\Omega [(-\Theta_x^2+\Theta_y^2) \sin
2\Theta+\nonumber\\
&&2\Theta_x\Theta_y\cos 2\Theta]\varphi_idA +\nonumber\\
&&\oint_\Gamma\sigma^\prime(\vartheta-\Theta)\varphi_ids\label{Vectorb}
\end{eqnarray}
we are able to solve for ${\bf \Theta}$ iteratively using a standard
numerical algorithm for the solution of systems of linear equations.
We have adopted the method of LU decomposition \cite{recipes} for
solving ${\bf \Theta}$.

\begin{figure}
\centerline{\epsfig{file=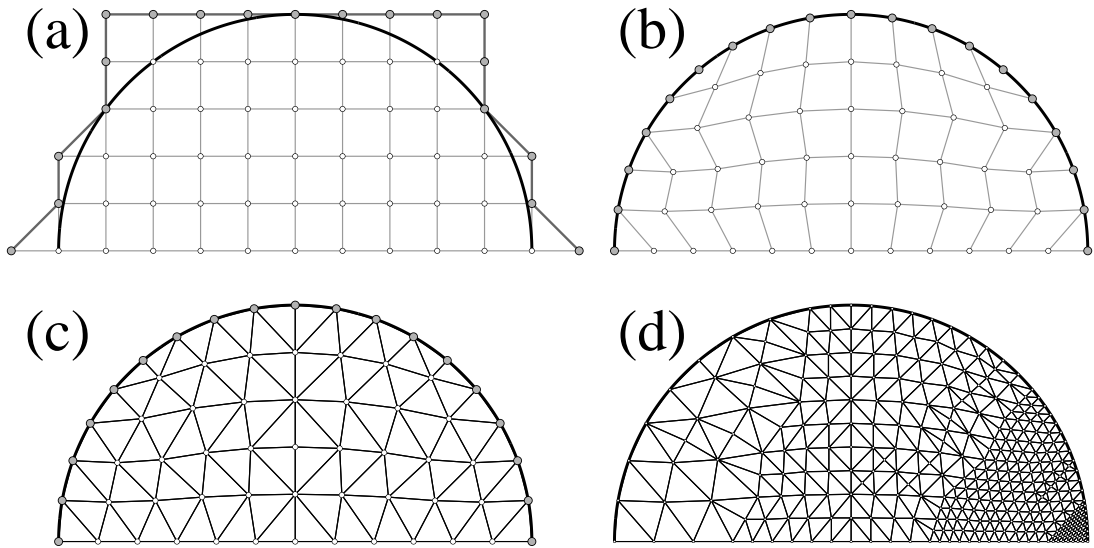, width=6in}}
\caption{The progressive steps towards mesh generation of the problem
for the domain, (a) initial choice of grid points in a square lattice,
(b) deforming to grid points into the region of interest $\Omega$ and
the bounding curve $\Gamma$ while keeping the square lattice topology
and connectivity, (c) triangulation using the square lattice connectivity
and (d) resulting mesh after adaptive mesh refinement.}
\label{mesht}
\end{figure}

The mesh generation algorithm plays an important role in the
efficiency of the FEM. An adaptive mesh generation algorithm is used
in our program to determine ${\bf \Theta}$.  We start with a mesh that
is nearly regular throughout $\Omega$ with a predefined grid size.
After obtaining a first estimate of ${\bf \Theta}$, a refined mesh is
generated.  The refined mesh has variable grid sizes over $\Omega$
depending on the variation of ${\bf \Theta}$.  Figure~\ref{mesht}
depicts the process of mesh generation with adaptive refinement.  We are
able to determine not only ${\bf \Theta}$, but also the derivatives
$\Theta_t$ and $\Theta_n$, which are necessary for the evaluation the
bounding curve, accurately and efficiently with the adaptive mesh
generation algorithm.

The next part of the algorithm is the determination of the bounding
curve $\Gamma$.  We assume the order parameter field $\Theta$ is fixed
in Eq.~(\ref{mGamma}) so as to simplify the problem.  We then pick an
origin in $\Omega$ and parameterize the bounding curve $\Gamma$ as
$k(\varphi)$, where $e^{k(\varphi)}\equiv|{\bf r}|$ is the distance
between the origin and $(x, y)$ on $\Gamma$, and $\varphi$ is the
polar angle.  In this parameterization, Eq.  (\ref{mGamma}) is a
second order non-linear differential equation in $k(\varphi)$.  If we
rewrite Eq.  (\ref{mGamma}) as
\begin{equation}
k^{\prime\prime} + q(\varphi; k, k^\prime)k^\prime=r(\varphi;k,
k^\prime)\label{RK}
\end{equation}
where $k^\prime\equiv dk/d\varphi$ and
\begin{eqnarray}
q(\varphi;k, k^\prime)&=&-\frac{\sigma^\prime\Theta_\varphi-
\sigma^{\prime\prime}\Theta_k}{\sigma+\sigma^{\prime\prime}}(1+k^{\prime2}),
\label{geq1}\\
r(\varphi;k, k^\prime)&=&\left[1-\frac{\sigma^\prime\Theta_k+
\sigma^{\prime\prime}\Theta_\varphi}{\sigma+\sigma^{\prime\prime}}\right]
(1+k^{\prime2})+\nonumber\\
&&\frac{{\mbox e}^k{\cal
H}_b+\lambda}{\sigma+\sigma^{\prime\prime}}(1+k^{\prime2})^{3/2}.
\label{geq2}
\end{eqnarray}
Again, it is possible to integrate the equation for $k(\varphi)$
iteratively using standard method for the solution of ordinary
differential equation.  The Runge-Kutta method \cite{recipes} is
chosen for our application.

The problem of solving Eqs.  (\ref{mbulk}), (\ref{mbc}) and
(\ref{mGamma}) for $\Theta(x, y)$ and $\Gamma$ is reformulated in
terms of the solution of Eqs.  (\ref{FEM}) and (\ref{RK}) iteratively
for ${\bf \Theta}$ and $\Gamma$.  We begin by assuming an initial
boundary $\Gamma^{(0)}$ and texture ${\bf \Theta}^{(0)}$, from which
the texture ${\bf\Theta}^{\prime(1)}$ can be computed using FEM. Then
iterated texture ${\bf\Theta}^{(1)}={\bf\Theta}^{(0)}+
({\bf\Theta}^{\prime(1)}-{\bf\Theta}^{(0)})/\nu_T^{(0)}$ is in turn
used to evaluate a new accepted boundary
$\Gamma^{(1)}=\Gamma^{(0)}+(\Gamma^{\prime(1)}-
\Gamma^{(0)})/\nu_B^{(0)}$, where $\Gamma^{\prime(1)}$ is obtained
using the Runge-Kutta ordinary differential equation integrator on Eq.
(\ref{mGamma}).  The process is repeated until both $\Delta{\bf
\Theta}^{(n)}\equiv|{\bf \Theta}^{(n)}-{\bf\Theta}^{(n-1)}|$ and
$\Delta\Gamma^{(n)}\equiv|\Gamma^{(n)}-\Gamma^{(n-1)}|$ are less then
a preset tolerance of the order $O(10^{-5})$.  The factors
$\nu_T^{(n)}$ and $\nu_B^{(n)}$ with initial magnitude in the order of
$O(10)$ is introduced to avoid numerical instability in these
iterative processes.  These factors, $\nu^{(n)}_T$ and $\nu^{(n)}_B$,
are adjusted depending on $\Delta{\bf \Theta}^{(n)}$
\end{multicols}
\widetext
\begin{multicols}{2}
\hspace{-.17in}
and $\Delta \Gamma^{(n)}$, 
and at the final iterations take on values close to unity.

\section{Domains}
\label{sec:domains}

It is well established that the boundary is strictly circular for a
domain with a boojum texture when elastic anisotropy and line-tension
anisotropy are not present, or $\gamma=b=0$~\cite{RudBru}.  This
texture-boundary combination is indeed a local minimum of
Eq.~(\ref{sysenr})\cite{PettyLuben}.  Thus, in order for there to be
non-circular domains, it is necessary to retain terms in the expansion
Eq.~(\ref{linetension}) up to at least terms going as $a_2$.  Using
the numerical algorithm described above, we have performed systematic
studies of the domain textures and shapes in terms of the elastic
anisotropy, the line-tension anisotropy as well as the domain size.
Before we describe our observation, we note that when $\gamma=b=0$,
the exact result is given by a circular boundary of radius $R_0$
together with a ``boojum'' texture with a $+2$ defect located a
distance $R_B\equiv R_0\left(1+\sqrt{1+\rho_0^2}\right)/\rho_0$ from
the center of the domain, where $\rho_0\equiv R_0a_1/\kappa$ is the
normalized domain radius~\cite{RudBru}.  An exact equilibrium
texture-boundary combination is shown in Fig.~\ref{exact}.  The
simulated image obtained using the Brewster angle microscopy (BAM) is
also displayed in the background.  The signature of a ``boojum'' in a
BAM image is a set of straight constant-intensity lines emerging from
a virtual defect slightly outside the domain.  The light intensity in
a BAM image depends on the exact experimental setup and the properties
of the monolayer~\cite{Teer}.  In the case of all simulated BAM images
presented in Fig.~\ref{exact} and elsewhere in this report, the
Brewster angle is taken to be that of water $\Theta_B=53.12^\circ$,
\vadjust{\vskip .1in 
\narrowtext
\begin{figure}
\centerline{\epsfig{file=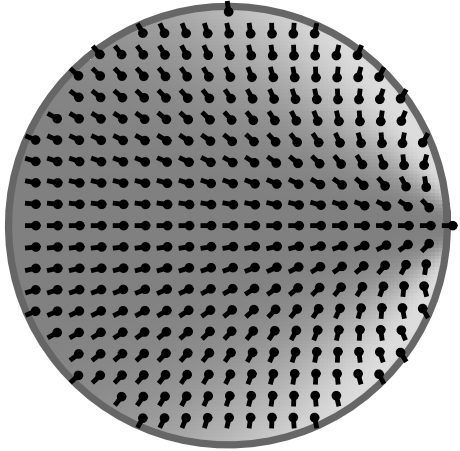, width=2.5in}}
\caption{Circular domain with a ``boojum'' texture, $R_0=5$, $\sigma_0=4$,
$a_1=1.6$, $a_2=0$, $\kappa=1$ and $b=0$. Also shown on the background is
the simulated image that would be obtained by Brewster angle microscopy.}
\label{exact}
\end{figure}
\vskip .05in}
the angle of the analyzer $\alpha$ is equal to $90^\circ$, the
thickness of the monolayer is assumed to be $d=0.3nm$, the tilt $\Psi$
is $30^\circ$, the dielectric constants of the monolayer are
$\epsilon_\perp=2.31$, $\epsilon_\parallel=2.53$ and it is assumed
that the wavelength of the light $\lambda=514nm$.

We first concentrate on the effects of $b$ and keep $\gamma=0$.  When
$b<0$, the texture is altered in such a way that the virtual defect
appears to move closer to the boundary.  This is observed as
accelerated convergence of the constant-intensity lines to a point on
the boundary.  On the other hand, when $b>0$, the texture relaxes as if
the virtual defect has moved away from the boundary.  The deviation of
the texture from the boojum texture is as large as $20\%$ when
$|b|\approx0.8$.  The textural response is qualitatively in accord
with that reported in Ref.~\cite{GalaFour}.  Although there are
significant textural corrections due to the presence of bulk elastic
anisotropy, the resultant textures very much resemble a ``boojum'' as
seen in a BAM image as shown in Fig.~\ref{bdepd}.  This means that it
is difficult to identify elastic anisotropy based on observation of
the textures in the domains.  The response of the boundary to
elastic anisotropy that we obtain contrasts to that reported in
Ref.~\cite{GalaFour}.  The domain acquires an indentation when $b<0$.
The indentation remains observable for a large range of domain sizes.
The boundary protrudes slightly for $b>0$, as depicted in
Fig.~\ref{bdepd}.  The protrusion when $b=+0.8$ is subtle and does not
resemble the sharp feature observed experimentally.  Thus, elastic
anisotropy 
\vadjust{\vskip .1in 
\narrowtext
\begin{figure}
\centerline{\epsfig{file=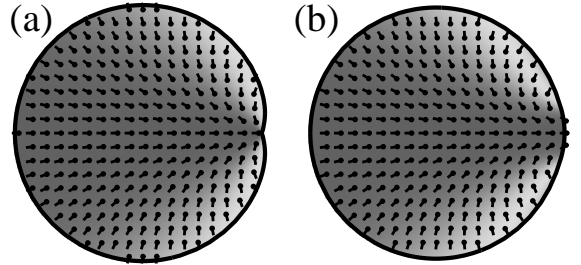, width=3in}}
\caption{The textures and the shapes of
domains with $R_0=5$, $\sigma_0=4$, $a_1=1.6$ and $a_2=0$.
Their stiffness coefficients are (a)$\beta=-0.8$, (b)$\beta=0.8$.}
\label{bdepd}
\end{figure}
\vskip .05in}
\vadjust{\vskip -.1in 
\narrowtext
\begin{figure}
\centerline{\epsfig{file=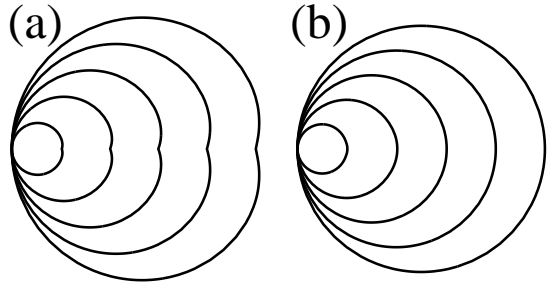, width=3in}}
\caption{The domain shapes computed for $a_0=4$, $a_1=1.6$ and $a_2=0$,
and $R_0=0.5$, $1$, $2$, $4$, $8$.
Their stiffness coefficients are (a)$\beta=-0.8$, (b)$\beta=0.8$.  For
ease of observation, domains are not shown to scale.}
\label{brdep}
\end{figure}
\vskip .05in}
\vadjust{\vskip .1in 
\narrowtext
\begin{figure}
\centerline{\epsfig{file=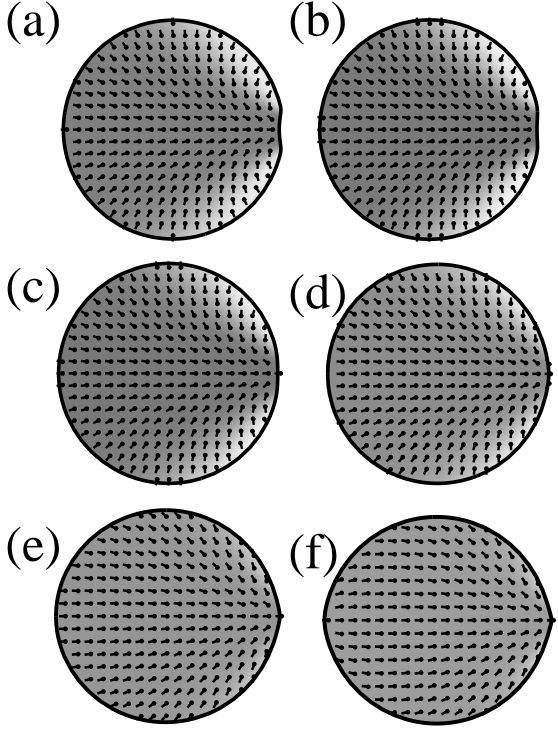, width=3in}}
\caption{The textures and shapes of domains computed for $-0.5<a_2<0.5$,
$R_0=5$, $\sigma_0=4$, $a_1=1$.  (a)$a_2=-0.5$, (b)$a_2=-0.3$, (c)
$a_2=-0.1$,( d) $a_2=0.1$,(e) $a_2=0.3$ and (f) $a_2=0.5$.}
\label{adepn}
\end{figure}
\vskip .05in}
alone is not capable of accounting for the shapes of the domains observed experimentally.  Figure~\ref{brdep} shows domains of
various sizes when (a) $b=-0.8$ and (b) $b=+0.8$.

We now proceed to discuss the role of the line-tension anisotropy,
parameterized by $\gamma$, in the textures and the boundaries of the
domains.  Elastic anisotropy will be eliminated ($b=0$) for simplicity.
We first investigate situations when $|\gamma| \leq1$.  For very small
domains where $R_0a_1/\kappa\ll1$, the texture is almost constant and
the dominant contribution to the boundary deformation comes from the
$a_2$ contribution.  The domain is elongated at both ends along the
axis connecting its center and the virtual defect when $\gamma>0$ and
is flattened at both ends along the same axis when $\gamma<0$.  Domain
shapes exhibit a 2-fold symmetry.  When $R_0a_1/\kappa \geq1$, the
texture closely resembles the boojum texture and contributes
significantly to the boundary distortion through the influence of
$\gamma$ in the line-tension.  The domain nolonger displays 2-fold
symmetry and acquires a protrusion when $\gamma>0$, or an indentation
when $\gamma<0$.  Figure \ref{adepn} shows domains of with $\gamma$
ranging from $-0.5$ to $0.5$.  The numerical algorithm also allows us
to examine domain shape and texture when $a_1=0$ and $a_2=1$.  In this
case, the domain acquires a ``cigar-shape'' and the texture is
associated with two virtual $+1$ defects~\cite{RudBru,pertb}.  The
progressive changes of the texture and the shape from a system with
$a_1=1$ and $a_2=0$ to one for which $a_1=0$ and $a_2=1$ are shown in
\vadjust{\vskip .1in 
\narrowtext
\begin{figure}
\centerline{\epsfig{file=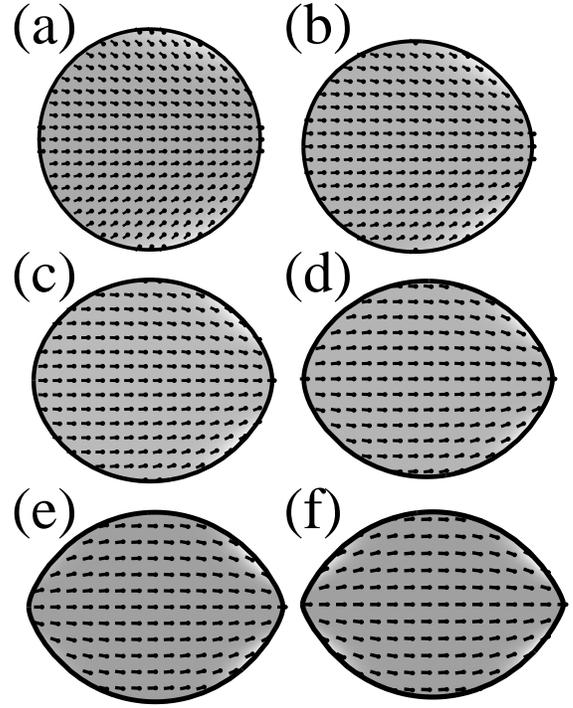, width=3.in}}
\caption{The textures and the shapes of domains
with $\rho=1$ and $a_0=4$.  Their coefficients of the anisotropy
line tensions are (a)$a_1=1$, $a_2=0$, (b)$a_1=0.8$, $a_2=0.2$,(c)$a_1=0.6$,
$a_2=0.4$,(d)$a_1=0.4$, $a_2=0.6$,(e)$a_1=0.2$, $a_2=0.8$ and
(f)$a_1=0$, $a_2=1$.}
\label{cgshpdn}
\end{figure}
\vskip .05in}
Fig.~\ref{cgshpdn}.  When both $a_1$ and $a_2$ are nonzero, the
texture can be thought of a superposition of pure $a_1$ and pure $a_2$
textures.  Typically at $R_0a_1/\kappa\sim 1$, the effect of the set
of two $+1$ defects become observable when $\gamma=1/4$.  Domains with
indentation, protrusions, and the ``cigar-shaped'' domains, have all
been observed experimentally~\cite{cigar}.

We have already briefly discussed the issue of size dependence in the
previous paragraph.  To look into this matter in detail, we will
examine the particular set of data reported in Ref.~\cite{FangTeer} in
which the domains investigated possess protruding features sharp
enough that ``excluded angles'', $\Psi_0$, characterizing the
boundaries can be identified.  The definition of $\Psi_0$ and the
experimental data are depicted in Fig.~\ref{andenf}.  The key features
of this set of experimental data are: (1) $\Psi_0$ goes through a
maximum as $R_0$ varies; (2) there is an abrupt onset of $\Psi_0$ in
the small $R_0$ region; and (3) the intercept at the $\Psi_0$-axis
when the curve is extrapolated implies
$\displaystyle{\lim_{R_0\rightarrow\infty}\Psi_0\neq0}$.

Before we make comparisons between theoretical results and the
experimental data, we comment on the extraction of $\Psi_0$ from
computed domain boundaries.  It has been shown that within the
parameter regime of our discussions, the domain boundaries are smooth
and continuous.  There is no cusp-like singularity on the boundary.
This can be seen in the domains of various sizes shown in
Fig.~\ref{nurdp}.  Nevertheless, $\Psi_0$ can be unambiguously
measured for some of these domains.  The values of the parameters
utilized here are $\kappa=1$, $\delta=0.4$ and $\gamma=0.5$.  To
determine $\Psi_0$ for these domains, we adopt a systematic scheme
that utilizes the function $I\equiv I_0\mbox{exp}[-(d^2x/dy^2)^2]$ to
capture the most likely $\Psi_0$ for a given domain bounding curve
devised in Ref.~\cite{pertb}, where $x(y)$ parameterizes $\Gamma$ in
Cartesian coordinates system.  Density plots of $I$ as a function of
$\Psi\equiv-2\tan^{-1}dx/dy$ and $R_0$ for numerical and the
perturbative results are shown, respectively, in Figs.~\ref{ddenna}(a)
and (c), the darker regions representing larger $I$, and highlighting
the more likely values of $\Psi_0$.  With the use of this method for
the determination of $\Psi_0$, we have obtained reasonable agreement
between the perturbative analysis and the numerical computations in
the large-$R_0$ regime.  We note here that the value at which $\gamma$
is set, 0.5, is too large for perturbative results to be dependable.
However, the perturbative results resemble those obtained numerically
in the sense that $\Psi_0$ increases as $R_0$ decreses from $\infty$.
The abrupt onset of $\Psi_0$ indicated in Figs.~\ref{ddenna}(c) and
(d) is not present in Figs.~\ref{ddenna}(a) and (b).  It is, however,
evident in Figs.~\ref{nurdp} that $\Psi_0$ can be unambiguously
identified for domains with $R_0\ge1$ while the domains become
elliptical for which $\Psi_0=0$ when $R_0<1$.  Hence, there is an
apparent jump in $\Psi_0$ near $R_0=1$ and beyond which $\Psi_0$
\vadjust{\vskip .1in 
\narrowtext
\begin{figure}
\centerline{\epsfig{file=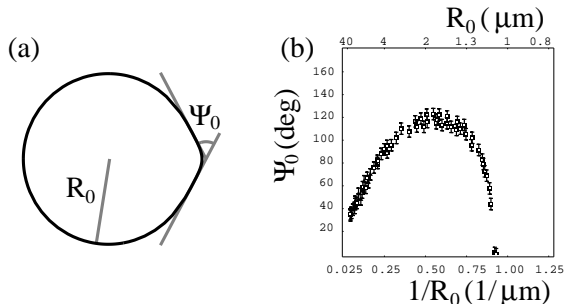, width=3in}}
\caption{(a) The definition of the excluded angle $\Psi_0$. (b)
Experimental measurements of the domain-size $R_0$ dependence of $\Psi_0$
observed in $L_2$ domains surrounded by LE phase taken from Ref. [9].}
\label{andenf}
\end{figure}
\vskip .05in}
\vadjust{\vskip -.2in 
\narrowtext
\begin{figure}
\centerline{\epsfig{file=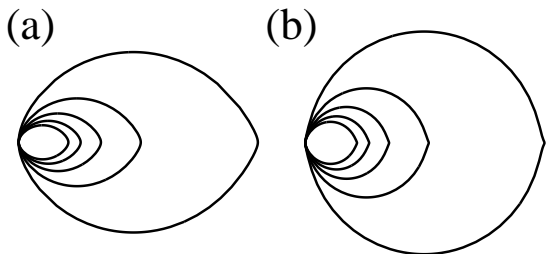, width=3in}}
\caption{The shapes of domains of various sizes computed for $\sigma_0=4$,
$a_1=1.6$ and $a_2=0.8$. (a)Smaller domains with $R_0=0.2$, $0.25$, $0.33$,
$0.5$, $1$
which exhibit 2-fold symmetry.  (b)Larger domains with $R_0=2$, $2.5$,
$3.3$, $5$, $10$
which have a protrusion on one end of the boundary.  Each of the sets of
domains are plotted to scale.}
\label{nurdp}
\end{figure}
\vskip .05in}
becomes non-zero.  The jump in $\Psi_0$ predicted in the perturbative
analysis\cite{pertb} is indeed confirmed by the more reliable
numerical computations reported here.  For very small domains
$(R_0<1)$, the shapes are predicted to be elliptical by our numerical
analysis, in contrast to the prediction of nearly circular domains
that results from the perturbative analysis.  The magnitude of
$\gamma$ that results in breakdown of the first order perturbative
analysis is the key origin of the mismatch.  In
figures~\ref{ddenna}(b) and (d), the maximum $I_{max}$ of $I$ is shown
as the dark line segments and the grey bands mark the regions in which
$I > I_{max}/2$.  They depict, respectively, numerical and
perturbative results.  Superimposed are the experimental data which
provides a reference for the comparisons described above.

To compare the theoretical results to the experimental data, a length
scale is required.  The length scale is set by the assignment
$\kappa/a_1=4\mu m$ when the comparisons is made between the
perturbative results and the experimental data~\cite{pertb}.  Except
for examining the results of more reliable computations, there is no
attempt to fit the experimental data in this report for reasons to be
discussed below.  We first adopt the same set of parameters, with
which the perturbative analysis fits the data in the large-$R_0$
regime, for the comparison.  It is obvious from Fig.~\ref{ddenna}(b)
that even at $\gamma=0.5$, the theoretical
\vadjust{\vskip .2in 
\narrowtext
\begin{figure}
\centerline{\epsfig{file=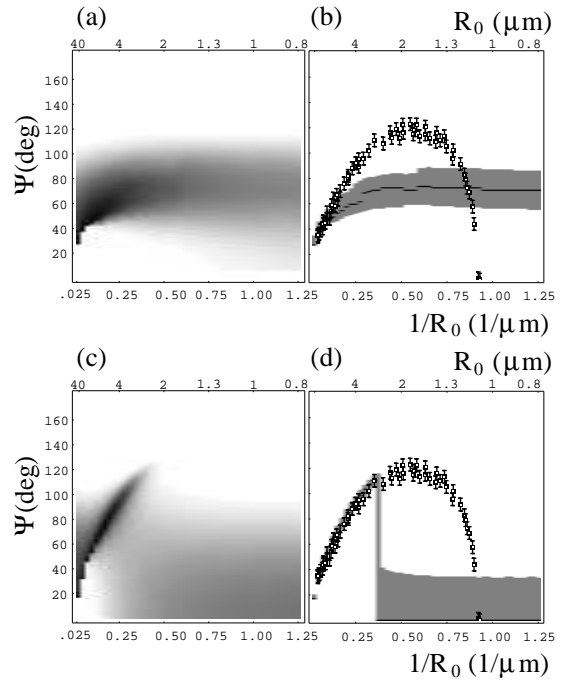, width=3in}}
\caption{(a) Density plot of $I$ as a function of $\Psi$ and $R_0$ of
the numerical results for the domains. (b) Plot of $I_{max}$ and the region in
which $I>I_{max}/2$ as a function of $\Psi$ and $R_0$ of the numerical results
for the domains.  Superimposed are the experimental data shown in
Fig.~\ref{andenf}(b) with parameters $\kappa/a_1=4\mu m$, $\delta=0.4$
and $\gamma=0.5$. (c) and (d) are corresponding plots for the perturbative
results.}
\label{ddenna}
\end{figure}
\vskip .05in}
prediction for the maximum
of $\Psi_0$ is much smaller than that observed experimentally.  This
superficial comparison between the maxima of $\Psi_0$ implies that
$\gamma$ is very much larger in the system investigated.  Detailed
comparisons do show excellent agreement for domains larger than $10\mu
m$.  Experimentally observed domains with maximum $\Psi_0$ and small
circular domains are not reproduced numerically.  Attempts have been
made to investigate the combined effect of elastic anisotropy $b$ and
$\gamma$.  However, for $\gamma$ of such a magnitude, contributions
from the $b$ do not affect the qualitative behaviors discussed in this
context.  It is 
thus concluded that although the simple elastic modelis not capable of fully addressing the issue of the domain size
dependence of the shapes, it has successfully produced the qualitative
features in the $\Psi_0$ versus $R_0$ plot and many nontrivial domain
shapes observed in various experiments\cite{cigar}.

In the large-$R_0$ regime ($R_0\gg1$), the boundary corrections are
confined in a small portion of the boundary and the domains become
nearly circular.  Because of the rapid texture variations in the
immediate vicinity of the boundary, associated with the approach to
the boundary of the virtual defect, we are unable to perform
dependable numerical investigations of extremely large domains.  This
leaves open the question of the asymptotic behavior of $\Psi_0$ is the
$R_0\rightarrow \infty$ limit.

With the numerical scheme for evaluating simultaneously ${\bf \Theta}$
and $\Gamma$.  We are able to explore the simple model
Eq.~(\ref{sysenr}) in a much wider range of the parameter space with

confidence.  Not only does the model account for the domains with
various features observed in experiments, it also yields an
appropriate domain size dependence of the boundary shapes.  However,
we are unable to perform reliable numerical investigations on
extremely large domains.  Despite the fact that there is an upper
bound to the domain size that we are able to compute, we believe, on
the basis of measurements of the defect positions \cite{RivMeu} that
the largest domains we are able to compute are not smaller than those
that have been observed experimentally.  The numerical algorithm
appears to be capable of evaluating domain shapes for arbitrary
anisotropic line-tension, with one caveat.  A closer look at Eqs.
(\ref{geq1}) and (\ref{geq2}) immediately indicates that this approach
\vadjust{\vskip .1in 
\narrowtext
\begin{figure}
\centerline{\epsfig{file=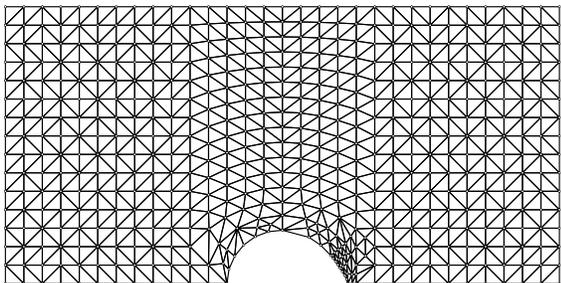, width=3in}}
\caption{Mesh with additional external boundary used in the direct
extension of the FEM algorithm for the bubbles.}
\label{dirbb}
\end{figure}
\vskip .05in}is not appropriate for situations in which
$\sigma+\sigma^{\prime\prime}=0$ at some points on the boundary.  An
approach that is appropriate to this situation is the Wulff
constructions~\cite{RudBru}.

\section{bubbles}
\label{sec:bubbles}

We now turn to the investigation of bubbles.  The first task is to
numerically evaluate the texture in a region, $\Omega$, that does not
have an external boundary.  It is possible to implement a
straightforward extension of the problem of the domain by introducing
an artificial external boundary far away from the inner bounding curve
$\Gamma$.  One must introduce a boundary condition on this added
external boundary by hand.  Figure ~\ref{dirbb} displays the
triangulations associated with such implementation of the approach to
the calculation of the property of a bubble.  The method, though
inefficient, produces results that are consistent with those obtained
perturbatively~\cite{pertb}.

The problem that involves an \emph{infinite} $\Omega$ with internal
boundary $\Gamma$ is referred to as the exterior problem.  If one does
not introduce an artificial external boundary, it is necessary to have
\vadjust{\vskip .1in 
\narrowtext
\begin{figure}
\centerline{\epsfig{file=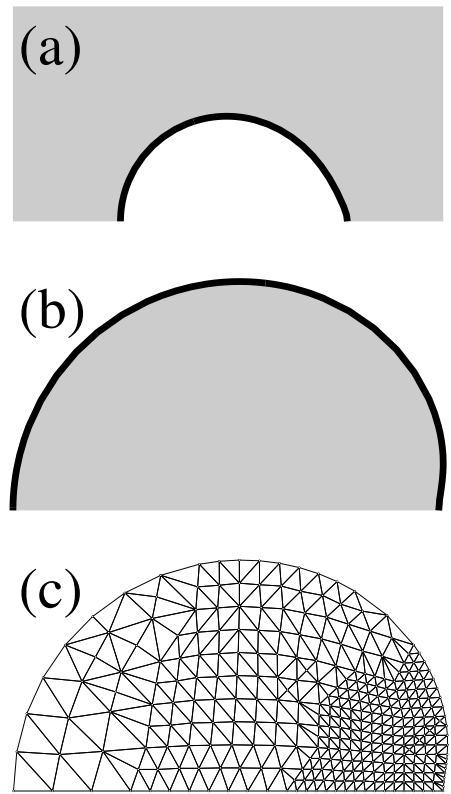, width=2.4in}}
\caption{Solving the problem of the bubble by transforming it into an
inverse domain problem. (a) shows the region $\Omega$ of ordered phase
and the boundary $\Gamma$ for the bubble. (b) shows the
transformed region $\Omega^\prime$ and boundary $\Gamma^\prime$. (c)
shows the mesh that covers $\Omega^\prime$.}
\label{trans}
\end{figure}
\vskip .05in}
\vadjust{\vskip .1in 
\narrowtext
\begin{figure}
\centerline{\epsfig{file=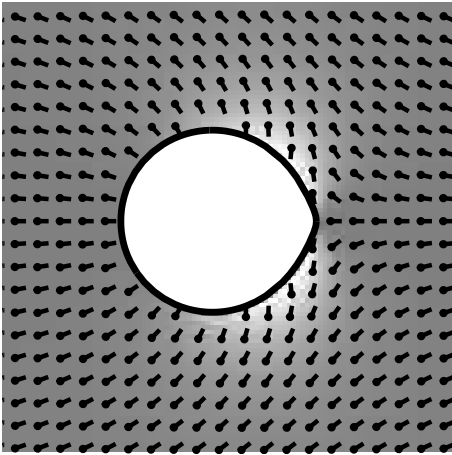, width=2.5in}}
\caption{Boundaries and textures around the bubbles for $R_0=5$, $a_1=1$,
$\kappa=1$ and $b=0$.}
\label{bbshp}
\end{figure}
\vskip .05in}
at hand a complete set of exterior solutions to construct the boundary
condition at $\Gamma$.  For our particular case, this is possible when
$b=0$, in which case the bulk extremum equation reduces to Laplace's
equation.  The examination of the problem with nonzero $b$ is a major
goal of this investigation, and we are not aware of the existence of
an appropriate set of external solutions in this case.  Noting that
the order parameter tends to a fixed value ($\Theta = 0$ for our case
here) as $r\rightarrow\infty$, it is possible to approach the problem 
of the bubble using a different set of polar co-ordinates, {\em i.e.} 
$(r^\prime, \varphi)\equiv (1/r, \varphi)$, that transform the bubble 
into a domain of area $\Omega^\prime$ and bounding curve $\Gamma^\prime$,
shown in Fig.~\ref{trans}, with the following ``elastic energy'',
\begin{eqnarray}
{\cal
H}=&&\frac{\kappa}{2}\int_{\Omega^\prime}\left\{\Theta_x^2+\Theta_y^2+
\beta\left[\left(-\Theta_x^2+\Theta_y^2\right)\cos 2(\Theta-2\varphi)\right.
 \right.  \nonumber\\
&&\left.\left.+2\Theta_x\Theta_y\sin 2(\Theta-2\varphi)\right]\right\}\,dA+
\oint_{\Gamma^\prime}
\frac{\sigma(\vartheta-\Theta)}{R^{-2}}\;ds\nonumber
\end{eqnarray}
With the problem transformed, the meshing algorithm used for the
domain can be applied immediately.  We are then able to proceed with
the investigation of the bubbles with the same efficiency and same
accuracy as the studies of the domains.

In the numerical studies that we have performed with the use of the
transformation above, the results for the bubbles reported in the
perturbative analysis~\cite{pertb,FangTeer} that the boundaries are
not circular even when $b=0$ and $a_2=0$, have been confirmed.
Figure~\ref{bbshp} shows the texture and the boundary of a typical
bubble.  In the background simulated BAM image, one notes the circular
constant-intensity lines which identify the ``inverse boojum''.  The
numerical algorithm further enables us to obtained equilibrium bubble
boundaries and textures around them when elastic anisotropies are
present.  As can be seen in Fig.~\ref{bdepb}, the elastic anisotropy
leaves the boundaries substantially unaffected while significantly
\vadjust{\vskip .1in 
\narrowtext
\begin{figure}
\centerline{\epsfig{file=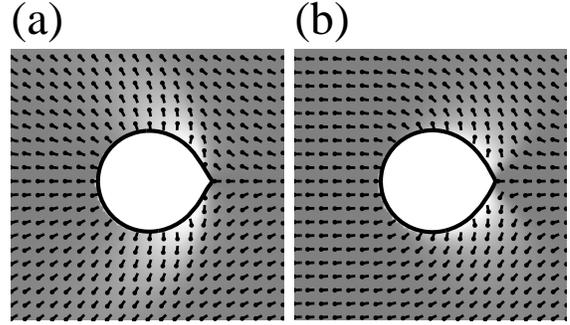, width=3in}}
\caption{Boundaries and textures around the bubbles for $R_0=5$, $a_1=1$,
$\kappa=1$,  (a) with $b=-0.8$ and  (b) $b=0.8$.}
\label{bdepb}
\end{figure}
\vskip .05in}
\vadjust{\vskip -.1in 
\narrowtext
\begin{figure}
\centerline{\epsfig{file=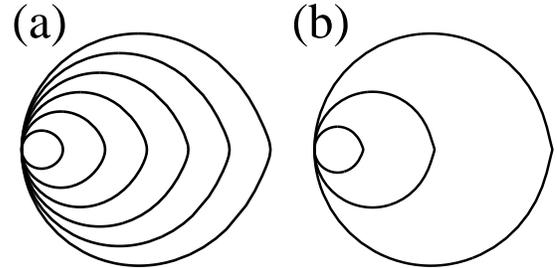, width=3in}}
\caption{Boundaries of bubbles computed for $\kappa=0.16$, $\sigma_0=1$,
$a_1=0.16$ and (a) $R_0=1$, $2$, $3$, $4$, $5$, $6$, and (b) $R_0=8$, $20$,
$40$.}
\label{bbrdpn}
\end{figure}
\vskip .05in}
changing the appearance of the textures around the bubbles.  The BAM
images are also shown in the same figures.  In contrast to the case in
which $b=0$, the constant-intensity lines become elongated
perpendicular to the axis connecting the center of the bubble and the
position of the virtual defect when $b<0$.  These lines are elongated
in the direction of the axis when $b > 0$, as shown.  This allows for
the determination of the sign of $b$ in the Langmuir monolayer by
examining the BAM images of the bubbles.

In Figs.~\ref{bbrdpn}, we display the size dependence of the bubble
boundaries.  Bubbles appear to be circular when they are small
($R_0<1$).  For large enough bubbles ($R_0\ge1$), an ``excluded
angle'' $\Psi_0$ defined in Fig.~\ref{andenf}(a) can be identified.
An approach similar to the analysis of the size dependence of the
boundary in the case of domains can be applied.
Figures~\ref{bdenna}(a) and (c) compare the density plots of $I$ as a
function of $\Psi$ versus $R_0$ for the numerical and the perturbative
results.  The two plots are in excellent agreement.  When $\gamma=0$,
the texture does not differ significantly from that of the inverse
boojum even though the bubble is not exactly circular.  This contrasts
to what is seen in domains when $\gamma\neq0$, in which case the
textures can deviate significantly from the boojum texture.  The $R_0$
dependence of $\Psi_0$ features that are qualitatively similar to
those seen in the case of the domains, i.e.~a maximum and an onset.  These
features have also been observed experimentally \cite{FangTeer}.
Experimental data is shown together with the numerical and
perturbative results in Figs.~\ref{bdenna}(b) and (d), respectively,
\vadjust{\vskip -.1in 
\narrowtext
\begin{figure}
\centerline{\epsfig{file=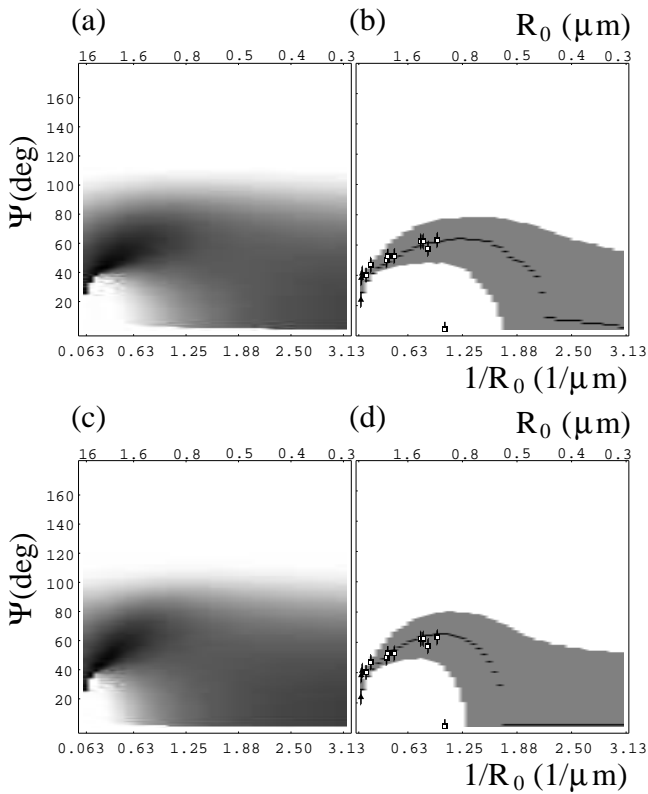, width=3.5in}}
\caption{(a) Plot of $I$ as a function of $\Psi$ and $R_0$ of the
numerical results for the bubbles.  (b) Plot
of $I_{max}$ and the region in which $I>I_{max}/2$ as a function of
$\Psi$ and $R_0$ of the numerical results for the bubbles.  Superimposed
are the experimental observations  of gaseous bubbles in $L_2$ phase.
The experimental data have appeared in Ref. [9].  The parameters for
the by-eye fit are $\kappa/a_1=0.4\mu m$ and $\delta=0.16$. (c) and
(d) are the corresponding plots for the perturbative results. }
\label{bdenna}
\end{figure}
\vskip .05in}
and all the results match reasonably well.  The by-eye fit
has been obtained in Ref.~\cite{FangTeer} and no further adjustment of
the parameters is made in this investigation.

We have thus devised a numerical method to approach the problem of the
bubble that can be implemented with the same efficiency as in the
problem of the domain.  There is good agreement between perturbative
and numerical results.  We are able to investigate the effect of the
elastic anisotropy, and our results point to a possible means for the
determination of the relative strength of $K_s$ and $K_b$.

\section{conclusions}
\label{sec:conclusions}
We have discussed in this report the implementation of a numerical
method that leads us to the solution of simultaneous equilibrium
conditions for the textures and the bounding curve of the domains.
Using this numerical algorithm, we have investigated the influences on
the textures and the domain shapes of the line-tension and elastic
anisotropies.  Our analysis of this simple model reveals that elastic
anisotropy does, indeed, result in interesting domain boundaries with
protrusions and indentations.  The domains with indentations resemble
those observed experimentally.  However, the domains that we generate with
protrusions are very different from those in observed in BAM images\cite{cigar}.
Hence, elastic anisotropy cannot qualitatively account for all
experimental observations.  Furthermore, our numerical results are in
contrast to the claims in Ref.~\cite{GalaFour}. Dents in
boundaries are due to a bend modulus that exceeds the splay modulus,
i.e.  $b\equiv(K_s-K_b)/(K_s+K_b)<0$, while protrusions are present
when $b>0$.  On the other hand, the second harmonic contribution to
the line-tension, parameterized by $\gamma\equiv a_2/a_1$ is capable
of producing nontrivial domain shapes that resemble the shapes
observed experimentally.  For the influence of $\gamma$ on the
boundary, our results are in qualitative agreement with those
presented in Ref.~\cite{GalaFour}.  Comparison has also been made
between perturbative results~\cite{pertb}, the numerical computations
described here and the experimental data\cite{FangTeer}.  The
magnitude of $\gamma$ used in the perturbative analysis is the prime
factor causing the mismatch between the perturbative and the numerical
results.  When $\gamma$ is large (=0.5 for our case), the first order
perturbative approach is not expected to be accurate.

While the results of the perturbative analysis and the numerical study
are different quantitatively, they possess similar
qualitative features, namely the onset of the excluded angle $\Psi_0$
as the domain size $R_0$ increases, and then $\Psi_0$ reaches a
maximum of and then decrease as $R_0$ continues to increase.
These match the qualitative features that are present the experimental
data \cite{FangTeer} shown in Fig.~\ref{andenf}(b).  Experimental
results is not reproduced in the numerical calculations when $R_0$ is
small.  The discrepancies between the experimental data and the
numerical result imply that other interactions, neglected in the
model, may be significant.

We have also extended the numerical algorithm to the problem of
bubbles.  It is found that the transformation $r\rightarrow
r^\prime\equiv 1/r$ results in a new domain problem which allows
us to solve the equilibrium conditions for the bubbles at the same
level of efficiency and accuracy as those for the domains.  Not only
have we obtained results that are consistent with those in the
perturbative analysis ~\cite{FangTeer,pertb}, we have also analyzed
the effect of elastic anisotropy on the textures and the boundary of
the bubbles, a task that is algebraically formidable in the
perturbative analysis.  The influence of elastic anisotropy on the
boundary is small while it significantly modifies the textures.  This
provides a means for the qualitative determination of the elastic
anisotropy by observing the texture around the bubbles.  The agreement
between the perturbative results and the numerical computation is
excellent.  This is not surprising as $a_2$ is not involved.  The
$R_0$ dependence of $\Psi_0$ is similar to the case of the domain,
except that the maximum of $\Psi_0$ is much smaller.  The perturbative
result agrees reasonably with the experimental data as reported in
Ref.~\cite{FangTeer}.  The numerical results match as well.

In conclusion, we have successfully implemented a numerical algorithm
that enable us to analyze unambiguously a simple model,
Eq.(\ref{sysenr}), of tilted ordered media in a non-trivial geometry
imposed by experimental observations.  Using this numerical algorithm
and its extensions, we are able to address the long-standing debate
with regard to the origin on the cusp-like features observed in
domains of Langmuir monolayers using an elastic model.  Within the
context of this simple model which addresses only the competition
between the bulk elastic energy and the boundary energy, many
qualitative features of the experimental observation have been
captured.  Discrepancies cannot be avoided, as the real system is much
more complex.  The model we adopted has neglected other effects and
interactions that are present in real system, such as dipolar
interactions and adjustments in the tilt degree of freedom.  A
combination of these effects may account for the discrepancies between
the experimental data and the theoretical results.  The apparently
general numerical algorithm is, however, not capable of handling
situation in which $\sigma+\sigma^{\prime\prime}=0$ at some points on
the boundary.  A different approach, such as the Wulff
construction\cite{RudBru}, is required.  Nevertheless, our numerical
algorithm is versatile and can be extended to systems containing
topological defects, or with the ordered phase filled in a nonsimply
connected space.

\section*{acknowledgments} We thank Professor Charles Knobler,
Professor Robijn Bruinsma, and Dr.  Jiyu Fang for useful discussions.
We are grateful to Professor Robert B. Meyer for his insightful
proposal of the compact and concise way of presenting the textures.

\appendix
\section{Variational formulation of the FEM}
In finite element analysis, we approximate $\Theta(x, y)$ by Eq.
(\ref{ThetaTria}).  The energy functional $H[\Theta]$ given in Eq.
(\ref{sysenr}) now becomes
\begin{eqnarray}
H({\bf \Theta})&=&\frac{\kappa}{2}\int_\Omega\left\{|\nabla\Theta|^2+b
\left[\left(-\Theta_x^2+\Theta_y^2\right)\cos
2\Theta-\right.\right.\nonumber\\
&&\left.\left.2\Theta_x\Theta_y\sin 2\Theta\right]\right\}\;dA
\oint_\Gamma\sigma(\vartheta-\Theta)\;ds\label{FEMHamil}
\end{eqnarray}
where we denote ${\bf\Theta}\equiv(\Theta_1, \Theta_2,
\cdots,\Theta_N)^T$, and $\Theta=\sum\varphi_i\Theta_i$.  We
differentiate Eq.  (\ref{FEMHamil}) with respect to a $\Theta_i$
yields
\begin{eqnarray}
\frac{\partial H({\bf\Theta})}{\partial\Theta_i}
&=&\kappa\int_\Omega[\varphi_{ix}
\varphi_{jx}(1-b\cos 2\Theta)+\nonumber\\
&&\varphi_{iy}\varphi_{jy}(1+b\cos
2\Theta)+\nonumber\\
&&b(\varphi_{ix}\varphi_{jy}+\varphi_{iy}\varphi_{jx})\sin 2\Theta]dA\Theta_j
\nonumber\\
&&-\kappa b\int_\Omega [(-\Theta_x^2+\Theta_y^2)\sin 2\Theta+\nonumber\\
&&2\Theta_x\Theta_y\cos 2\Theta]\varphi_idA -\oint_\Gamma\sigma^\prime(\vartheta-\Theta)\varphi_ids
\end{eqnarray}
The equilibrium condition gives ${\bf A}{\bf \Theta}={\bf b}$ with
${\bf A}$ and ${\bf b}$ provided in Eqs.  (\ref{MatrixA}) and
(\ref{Vectorb}).

\section{Integration over a triangulation}
The integrals in Eqs.  (\ref{MatrixA}) and (\ref{Vectorb}) over
$\Omega$ are broken up into sums of integration over the triangles in
the triangulation of $\Omega$.  Integration over the interior
individual triangle can usually be carried out analytically depending
on the specific forms of the basis functions $\varphi_i(x, y)$ and the
matrix elements $A_{ij}$ and $b_i$.  We have chosen $\varphi_i(x, y)$
to be a continuous, piecewise linear function in $x$ and $y$ within a
triangle.  The line integral $\oint_\Gamma ds$ in $b_i$ will must be
evaluated numerically, because integrand depends on the polar angle
$\varphi$, which is not linear in $x$ or $y$.  This does not degrade
the efficiency of the computation because first of all, only triangles
whose perimeters coincide with $\Gamma$ contribute to the line
integral and secondly it is a line integral over a short distance.

In Eq. (\ref{ThetaTria}),  we express a function $f(x, y)$ for $(x, y) \in
\Omega$ in terms of its values at the nodes of the triangulation and the
corresponding basis functions $\varphi_i(x, y)$.  Within an individual
triangle $K$,  we can write
\begin{eqnarray}
f(x, y)=\sum_{i=1}^{3}f_{K_i}\varphi^{(K)}_{K_i}(x, y)\label{restriction}
\end{eqnarray}
where $K_i$ is the index of the $i$th vertex of the triangle $K$,
$f_{K_i}= f(x_{K_i}, y_{K_i})$, $(x_{K_i}, y_{K_i})$ are respectively
the functional value of $f(x, y)$ and the coordinates of the $i$th
vertex.  $\varphi^{(K)}_{K_i}(x, y)$ is the restriction of
$\varphi_{K_i}(x, y)$ in $K$.
The actual index of the $i$th vertex is $K_i$.  It is, however,
awkward to carry the $K$ in the symbol $K_i$ throughout the
discussion.  We will use $i$ to identity the vertex for simplicity
from now on, i.e.  $f_{K_i}$ is simplified as $f_i$.  We introduce a
set of natural coordinates $u$ and $v$ such that
\begin{eqnarray}
f(u, v)=f_1+(f_2-f_1)u + (f_3-f_1)v, \label{universal}
\end{eqnarray}
where $u\in[0, 1]$, $v\in[0,1]$ and $u+v\leq1$.  Transformation
between variable sets $x-y$ and $u-v$ can be obtained from Eq.
(\ref{universal}) by substituting $f$ with $x$ and $y$.  We then have
the followings relations
\begin{eqnarray}
u=&&\frac{1}{\Delta}\left[\left(y_3-y_1\right)x-\left(x_3-x_1\right)y+
x_3y_1-x_1y_3\right],\\
v=&&\frac{1}{\Delta}\left[\left(y_1-y_2\right)x-\left(x_1-x_2\right)y+
x_1y_2-x_2y_1\right],
\end{eqnarray}
where $\Delta$ is the Jacobian determinant given by
\end{multicols}
\renewcommand{\thesection}{{\Alph{section}}}
\renewcommand{\theequation}{{\thesection.\arabic{equation}}}
\widetext
\begin{eqnarray}
\Delta=\frac{\partial(x, y)}{\partial(u, v)}=
x_1y_2-y_1x_2+x_2y_3-x_3y_2+x_3y_1-x_1y_3.
\end{eqnarray}
Identifying Eqs.  (\ref{restriction}) and (\ref{universal}), we find
$\varphi_1 (u,v)=1-u-v$, $\varphi_2(u, v)=u$ and $\varphi_3(u, v)=v$.
In terms of $x-y$, we have
\begin{eqnarray}
\varphi_1(x, y)
=&&\frac{1}{\Delta}\left[\left(y_2-y_3\right)x-\left(x_2-x_3\right)y+
x_2y_3-x_3y_2\right],\\
\varphi_2(x, y)=&&\frac{1}{\Delta}\left[\left(y_3-y_1\right)x-
\left(x_3-x_1\right)y+x_3y_1-x_1y_3\right],\\
\varphi_3(x, y)=&&\frac{1}{\Delta}\left[\left(y_1-y_2\right)x-
\left(x_1-x_2\right)y+x_1y_2-x_2y_1\right].
\end{eqnarray}

Evaluation of the matrix element $A_{ij}$ involves of the following
area integrals which can be computed analytically.  The trivial one is
the area of $K$, which is $\int_K\;dA=|\Delta|/2$ and
\begin{eqnarray}
\int_K \cos 2\Theta \;dA
=&&\frac{|\Delta|}{4}\left[\frac{\cos 2\Theta_1}{(\Theta_1-\Theta_2)
(\Theta_2-\Theta_3)}+\frac{\cos
2\Theta_2}{(\Theta_2-\Theta_3)(\Theta_3-\Theta_1)}+\frac{\cos
2\Theta_3}{(\Theta_3-\Theta_1)(\Theta_1-\Theta_2)}\right],\\
\int_K \sin 2\Theta \;dA
=&&\frac{|\Delta|}{4}\left[\frac{\sin 2\Theta_1}{(\Theta_1-\Theta_2)
(\Theta_2- \Theta_3)}+\frac{\sin
2\Theta_2}{(\Theta_2-\Theta_3)(\Theta_3-\Theta_1)}+\frac{\sin
2\Theta_3}{(\Theta_3-\Theta_1)(\Theta_1-\Theta_2)}\right],
\end{eqnarray}
as $\varphi_{ix}$ and $\varphi_{iy}$ are constants.  Evaluation of
$b_i$ involves
\begin{eqnarray}
\int_K\varphi_1\cos 2\Theta\;dA
=&&\frac{|\Delta|}{4}\left[\frac{\cos 2\Theta_1}{(\Theta_3-\Theta_1)(\Theta_1-
\Theta_2)}+\frac{\sin 2\Theta_1-\sin
2\Theta_3}{2(\Theta_1-\Theta_3)^2(\Theta_3-
\Theta_2)}-\frac{\sin 2\Theta_1-\sin
2\Theta_2}{2(\Theta_1-\Theta_2)^2(\Theta_3-
\Theta_2)}\right],\\
\int_K\varphi_2\cos 2\Theta\;dA
=&&\frac{|\Delta|}{4}\left[\frac{\cos 2\Theta_2}{(\Theta_1-\Theta_2)(\Theta_2-
\Theta_3)}+\frac{\sin 2\Theta_2-\sin
2\Theta_3}{2(\Theta_2-\Theta_3)^2(\Theta_3-
\Theta_1)}-\frac{\sin 2\Theta_2-\sin
2\Theta_1}{2(\Theta_1-\Theta_2)^2(\Theta_3-
\Theta_1)}\right],\\
\int_K\varphi_3\cos 2\Theta\;dA
=&&\frac{|\Delta|}{4}\left[\frac{\cos 2\Theta_3}{(\Theta_2-\Theta_3)(\Theta_3-
\Theta_1)}+\frac{\sin 2\Theta_3-\sin
2\Theta_2}{2(\Theta_2-\Theta_3)^2(\Theta_2-
\Theta_1)}-\frac{\sin 2\Theta_3-\sin
2\Theta_1}{2(\Theta_1-\Theta_3)^2(\Theta_3-
\Theta_1)}\right],\\
\int_K\varphi_1\sin 2\Theta\;dA
=&&\frac{|\Delta|}{4}\left[\frac{\sin 2\Theta_1}{(\Theta_3-\Theta_1)(\Theta_1-
\Theta_2)}-\frac{\cos 2\Theta_1-\cos
2\Theta_3}{2(\Theta_1-\Theta_3)^2(\Theta_3-
\Theta_2)}+\frac{\cos 2\Theta_1-\cos
2\Theta_2}{2(\Theta_1-\Theta_2)^2(\Theta_3-
\Theta_2)}\right],\\
\int_K\varphi_2\sin 2\Theta\;dA
=&&\frac{|\Delta|}{4}\left[\frac{\sin 2\Theta_2}{(\Theta_1-\Theta_2)(\Theta_2-
\Theta_3)}-\frac{\cos 2\Theta_2-\cos
2\Theta_3}{2(\Theta_2-\Theta_3)^2(\Theta_3-
\Theta_1)}+\frac{\cos 2\Theta_2-\cos
2\Theta_1}{2(\Theta_1-\Theta_2)^2(\Theta_3-
\Theta_1)}\right],\\
\int_K\varphi_3\sin 2\Theta\;dA
=&&\frac{|\Delta|}{4}\left[\frac{\sin
2\Theta_3}{(\Theta_2-\Theta_3)(\Theta_3-
\Theta_1)}-\frac{\cos 2\Theta_3-\cos
2\Theta_2}{2(\Theta_2-\Theta_3)^2(\Theta_2-
\Theta_1)}+\frac{\cos 2\Theta_3-\cos
2\Theta_1}{2(\Theta_1-\Theta_3)^2(\Theta_3-
\Theta_1)}\right].
\end{eqnarray}
The above formulae will work only if
$\Theta_1\neq\Theta_2\neq\Theta_3$.  Let us consider cases where the
values of $\Theta=\Theta_E$ at 2 vertices of triangle $K$ and
$\Theta=\Theta_O$ at the other vertex.  We obtain the following for
the integrals in $A_{ij}$
\begin{eqnarray}
\int_K\cos 2\Theta\;dA&=&\frac{|\Delta|}{4}\left[\frac{\cos 2\Theta_E-\cos
2\Theta_O}
{(\Theta_E-\Theta_O)^2}+\frac{2\sin 2\Theta_E}{\Theta_E-\Theta_O}\right]\\
\int_K\sin 2\Theta\;dA&=&\frac{|\Delta|}{4}\left[\frac{\sin 2\Theta_E-\sin
2\Theta_O}
{(\Theta_E-\Theta_O)^2}-\frac{2\cos 2\Theta_E}{\Theta_E-\Theta_O}\right]
\end{eqnarray}
We denote $\varphi_E$ the restrictions of the basis functions at the
nodes that have $\Theta=\Theta_E$, and $\varphi_O$ the restriction of
the basis function at the node that has $\Theta=\Theta_O$.  We arrive
at
\begin{eqnarray}
\int_K\varphi_E\cos 2\Theta\;dA&=&\frac{|\Delta|}{4}\left[\frac{\sin 2\Theta_E}
{\Theta_E-\Theta_O}+\frac{\cos 2\Theta_E}{(\Theta_E-\Theta_O)^2}-\frac{\sin 2
\Theta_E-\sin 2\Theta_O}{2(\Theta_E-\Theta_O)^3}\right]\\
\int_K\varphi_O\cos 2\Theta\;dA&=&\frac{|\Delta|}{4}\left[-\frac{\cos
2\Theta_E+
\cos 2\Theta_O}{(\Theta_E-\Theta_O)^2}
+\frac{\sin 2\Theta_E-\sin 2\Theta_O}{(\Theta_E-\Theta_O)^3}\right]\\
\int_K\varphi_E\sin 2\Theta\;dA&=&\frac{|\Delta|}{4}\left[-\frac{\cos
2\Theta_E}
{\Theta_E-\Theta_O}+\frac{\sin 2\Theta_E}{(\Theta_E-\Theta_O)^2}+\frac{\cos 2
\Theta_E-\cos 2\Theta_O}{2(\Theta_E-\Theta_O)^3}\right]\\
\int_K\varphi_O\sin 2\Theta\;dA&=&\frac{|\Delta|}{4}\left[-
\frac{\sin 2\Theta_E+\sin 2\Theta_O}{(\Theta_E-\Theta_O)^2}-\frac{\cos
2\Theta_E-
\cos 2\Theta_O}{(\Theta_E-\Theta_O)^3}\right]
\end{eqnarray}
for the integrals required to evaluate $b_i$. Finally,  when
$\Theta_i=\Theta_E$
for all $i$'s, one will need $\int_K\varphi_E\;dA=|\Delta|/6$.
\begin{multicols}{2}

\section{Derivatives on the boundary}
One of the biggest benefits of the FEM is that it enables
straightforward determination of the derivatives of $\Theta$ on the
boundary.  The tangential derivative at node $i$ as
\begin{eqnarray}
\left.\frac{\partial\Theta}{\partial t}\right|_i=\frac{1}{2e^k
\sqrt{1+k^{\prime2}}}\left[\frac{\Theta_{i+1}-\Theta_i}
{\varphi_{i+1}-\varphi_i}+\frac{\Theta_i-\Theta_{i-1}}{\varphi_i-
\varphi_{i-1}}\right]
\end{eqnarray}
The normal derivative of $\Theta$ is given by Eq. (\ref{mbc}) which reads
\begin{eqnarray}
\frac{\partial\Theta}{\partial
n}=\frac{e^k\sigma^\prime(\vartheta-\Theta)+ \kappa b \Theta_t\sin
2(\Theta-\vartheta)}{\kappa[1-b\cos 2(\Theta-\vartheta)]}.
\end{eqnarray}

\end{multicols}


\begin{references}
\bibitem{phases} C.M. Knobler and R.C. Desai, Annu. Rev. Phys. Chem.
{\bf 43}, 207 (1992).

\bibitem{stripe} J. Ruiz-Garcia, X. Qiu, M.-W. Tsao, G. Marshall, C.M.
Knobler, G. A. Overbeck and D. M\"{o}bius, J. Phys.  Chem {\bf 97},
6955 (1993); D.K. Schwartz, J.Ruiz-Garcia, X. Qiu, J.V. Seligner and
C.M. Knobler, Physica A {\bf 204}, 606 (1994);J.V. Selinger, Z.-G.
Wang, R.F. Bruinsma and C.M. Knobler, Phys.  Rev.  Lett.  {\bf 70},
1139 (1993).

\bibitem{star}
X. Qiu, J. Ruiz-Garcia, K. J. Stine, C.M. Knobler, J.V. Selinger, Phys. Rev.
Lett. {\bf 67} 703 (1991).

\bibitem{FiscBru} T.M. Fischer, R.F. Bruinsma and C.M. Knobler, Phys.
Rev.  E {\bf 50}, 413 (1994).

\bibitem{RivMeu} S. Rivi\`{e}re and J. Meunier, Phys. Rev. Lett. {\bf 74},
2495 (1995).

\bibitem{Mermin} N.D. Mermin, in {\em Quantum Fluids and Solids},
edited by S.B. Trickey, E. Adams and J. Duffy (Plenum, New York,
1977).

\bibitem{FangTeer} J. Fang, E. Teer, C.M. Knobler, K.-K. Loh and J.
Rudnick, Phys.  Rev.  E {\bf 56}, 1859 (1997).

\bibitem{LanSeth} S.A. Langer and J.P. Sethna, Phys.  Rev.  A {\bf
34}, 5035 (1986).

\bibitem{Kraus}
I. Kraus and R.B. Meyer, Phys. Rev. Letts {\bf 81}, 3815 (1999).

\bibitem{RudBru} J. Rudnick and R. Bruinsma, Phys. Rev. Lett. {\bf 74},
2491 (1995).

\bibitem{GalaFour} P. Galatola and J.B. Fournier, Phys. Rev. Lett. {\bf 75},
3297 (1995).

\bibitem{pertb} J. Rudnick and K.-K. Loh,  Phys. Rev. E {\bf 60}, 3045 (1999).

\bibitem{FEM} C. Johnson, {\em Numerical solution of partial
differential equations by the finite element method} (Cambridge
University Press, Cambridge, 1987).

\bibitem{recipes} W.H. Press, B.P. Flannery, S.A. Tuekolsky and
W.T. Vetterling, {\em Numerical Recipes in C, The Art of Scientific
Computing} (Cambridge University Press, Cambridge, 1991).

\bibitem{numerics} K.-K. Loh and J. Rudnick,   Phys. Rev. Lett. {\bf 81},
4935 (1998).

\bibitem{PettyLuben} D. Pettey and T.C. Lubensky, Phys. Rev. E
{\bf 59}, 1834 (1999).

\bibitem{Teer} E. Teer and C.M. Knobler, private communications;
C. Lautz, T.M. Fischer, M. Weygand, M. L\"{o}sche, P.B. Howes and K.
Kjaer, J. Chem. Phys. {\bf 108}, 4640 (1998).

\bibitem{cigar} J. Fang and C.M. Knobler, unpublished.

\end{references}
\end{document}